\documentstyle[12pt]{article}
\input{epsfig.sty}
\newcommand{\beq}{\begin{eqnarray}}
\newcommand{\eeq}{\end{eqnarray}}
\begin{document}
\title{$Z_c(3900)$ as a Four-Quark State}
\author{Leonard S. Kisslinger and Steven Casper\\
Department of Physics, Carnegie Mellon University, Pittsburgh, PA 15213}
\date{}
\maketitle

\begin{abstract}
   Using the method of QCD Sum Rules, we derive the correlator $\Pi^Z$ for a 
state consisting of two charm quarks and two light quarks, $c\bar{d}\bar{c}u$,
and carry out a Borel transform to find $\Pi^Z(M_B)$. From this we find the 
solution that  $M_B\simeq 3.9 \pm .2$ GeV, showing that the $Zc(3900)$ is a
tetra-quark state.

\end{abstract}
\maketitle
\noindent
PACS Indices:11.15.-q,12.38.Lg,14.40.Lb,14.65.Dw


\section{Introduction}

  Recently $e^+ e^-$ collision experiments by BESIII\cite{BESIII13} and
Belle\cite{Belle13} Collaborations have found a state at about 3,900 MeV,
called the $Z_c(3900)$\cite{BESIII13}, that might be a four quark state, 
$|c\bar{d}\bar{c}u>$\cite{es13}

  For many years states in the energy region of 3,900-4,500 MeV, near the
$D \bar{D}$ threshold, there have been found possible tetra-quark states.
See, e.g., Ref\cite{mprp05}, for a theoretical study of the X(3872) as
a charm tetra-quark about one decade ago. Recently, there have been studies
of the  $Z_c(3900)$ as a $\bar{D}D^*$ molecular state\cite{zhang13,wang14},
vector/axial vector Charmonium state\cite{chen11}
using the method of QCD Sum Rules as in the present study. See these 
references for references to earlier publications, as well as a review of
New Charmonium States via QCD Sum Rules\cite{nielsen10}

In our study of the $Z_c(3900)$ as possibly a $|c\bar{d}\bar{c}u>$ state we 
use the method of QCD sum rules\cite{svz79}. Our approach differers from
earlier studies\cite{zhang13,wang14} in that our correlator corresponds to a
four-quark (tetra-quark) rather than a $\bar{D}D^*$ molecular state.
First  we find the correlator in momentum space, and then
as a function of the Borel mass, $M_B$, and see if it has a 
minimum near the value of 3.9 GeV, similar to our study of heavy quark hybrid
meson states\cite{lsk09}.

  In Section II we briefly review the method of QCD sum rules, and derive the 
correlator for our four-quark model. In Section III we find the correlator as 
a function of the Borel mass in the region near 4,000 Mev; and then from a plot
find the minimum value of $M_B$. In Section IV we discuss the results and 
conclusions.
\newpage

\section{The $|c\bar{d}\bar{c}u>$ state and QCD Sum Rules}

  The method of QCD sum rules\cite{svz79} for finding the mass of a state A 
starts with the correlator,
 
\beq
\label{2}
       \Pi^A(x) &=&  \langle 0 | T[J_A(x) J_A(0)]|0 \rangle \; ,
\eeq
with $|0 \rangle$ the vacuum state and
the current $J_A(x)$ creating the states with quantum numbers A:
\beq
\label{3}
     J_A(x)|0 \rangle &=& c_A |A \rangle + \sum_n c_n |n; A  \rangle  \; ,
\eeq
where $ |A \rangle$ is the lowest energy state with quantum numbers A,
and the states $|n; A  \rangle$ are higher energy states with the A quantum
numbers, which we refer to as the continuum. One then carries out a Borel
transform to reduce the importance of the continuum and higher order diagrams.

  For our theory of the $Z_c(3900)$ as a  tetra-quark state
$|c\bar{d}\bar{c}u>$, we use the current $J_{Z_c}$, which creates a 
$J^{PC}=1^{+-}$ tetra-quark state:
\beq
\label{current}
       J_{Z_c}&=& \bar{d}\gamma_5 c \bar{c}\gamma_5 u \; ,
\eeq
with the correlator
\beq
\label{correlator}
       \Pi^Z(x)&=& \langle 0| T[J_{Z_c}(x) J_{Z_c}(0)]|0 \rangle \; .
\eeq

  Note that as was emphasized by J-R. Zhang in the Summary of Ref\cite{zhang13}
the current and correlator used for a QCD sum rule study of the $Z_c(3900)$ as 
a $\bar{D}D^*$ molecular state, with a current and correlator given by
\beq
\label{molecular}
    J^\mu_{\bar{D}D^*}&=&(\bar{Q}_a i\gamma_5 q_a)(\bar{q}_b\gamma^\mu Q_b)
\nonumber \\
 \Pi^{\mu \nu}_{\bar{D}D^*}&=& i <0|T[J^\mu_{\bar{D}D^*}(x)
J^{\nu +}_{\bar{D}D^*}(0)]|0> \; ,
\eeq
with $Q,q$ charm,light ($u,d$) quarks and $a,b$ color indices,
is just one possible theoretical interpretation of the $Z_c(3900)$ state,
and is not the same at a tetra-quark state. We emphasize that with our 
current, Eq(\ref{current}) and correlator, Eq(\ref{correlator}), we are 
using a QCD sum rule to explore the possibiity that the $Z_c(3900)$ is a 
tetra-quark state, which is not completely orthoganal to but is quite
different from a $\bar{D}D^*$ molecular state, by finding the mass using the 
tetra-quark correlator.

Using Wick's Theorem to express $\Pi^Z(x)$ in terms of the quark 
propagators, and taking the Fourier transform, one obtains the correlator 
in momentum space:
\beq
\label{correlatorp}
     \Pi^Z(p)&=& \int \frac{d^4k_1 d^4k_2 d^4k_3}{((2\pi)^4)^3}
Tr[S_d(k_1)\gamma_5 S_c(k_2)\gamma_5]Tr[S_c(k_3)\gamma_5 S_u(p+k_1-k_3+k_3)
\gamma_5] 
\; ,
\eeq
\newpage
with $S_q(k)$ a quark propagator,
\beq
\label{qprop}
       S_q(k)&=& \frac{\not{k} +m_q}{k^2-m_q^2} \; ,
\eeq
where $\not{k}=\sum_\alpha \gamma^\alpha k_\alpha$, with $\gamma^\alpha$ a
Dirac matrix.

The correlator is illustrated in Fig. 1
 
  Note that higher order terms are very small, as pointed out in 
Ref\cite{wang14}.

\begin{figure}[ht]
\begin{center}
\epsfig{file=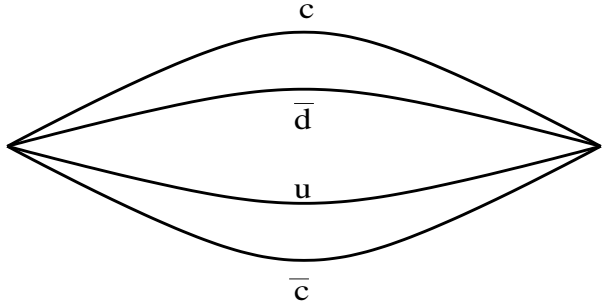,height=4cm,width=8cm}
\caption{$c,\bar{c}$ are charm, anticharm quarks. $u,\bar{d}$ are up, 
antidownquarks}
\label{Fig.1}
\end{center}
\end{figure}

Finally, we carry out a Borel Transform\cite{svz79}
\beq
\label{borel}
        {\cal B} \Pi^Z(p) &=&  \Pi^Z(M_B)
\eeq
\section{The Correlator $\Pi^Z(p) \rightarrow \Pi^Z(M_B)$}

From Eqs(\ref{correlatorp},\ref{qprop}), carrying out the traces, one finds
\beq
\label{PiZ(p)}
  \frac{\Pi^Z(p)}{16}&=&  \int \frac{d^4k_1 d^4k_2 d^4k_3}{((2\pi)^4)^3}
\frac{mM(mM-k_1\cdot k_2)+k_3\cdot(p+k_1-k_2+k_3)(k_1\cdot k_2-mM)}
{(k_1^2-m^2)(k_2^2-M^2)(k_3^2-M^2)((p+k_1-k_2+k_3)^2-m^2)} \; .
\eeq

  Since the $k_1\cdot k_2$ term vanishes via the momentum integrals and the
$k_3\cdot(p+k_1-k_2+k_3)$ term vanishes via the Borel transform, one needs to
evaluate two terms;
\beq
\label{I1}
       I_1(p)&=& 16 \int \frac{d^4k_1 d^4k_2 d^4k_3}{((2\pi)^4)^3}
\frac{m^2 M^2}{(k_1^2-m^2)(k_2^2-M^2)(k_3^2-M^2)((p+k_1-k_2+k_3)^2-m^2)} \; .
\eeq
\beq
\label{I4}
       I_4(p)&=& 16 \int \frac{d^4k_1 d^4k_2 d^4k_3}{((2\pi)^4)^3}
\frac{k_1\cdot k_2 k_3\cdot (p+k_1-k_2+k_3)}{(k_1^2-m^2)(k_2^2-M^2)
(k_3^2-M^2)((p+k_1-k_2+k_3)^2-m^2)} 
\; ,
\eeq
with $M=M({\rm charm\;quark})\simeq$ 1.5 GeV and $m=m(u,d {\rm\;quark})\simeq$
4 MeV.
\newpage

  For the evaluation of Eqs(\ref{I1},\ref{I4}) one uses
\beq
\label{IH}
     I_H(p)&=& \int \frac{d^4k}{(2 \pi)^4}\frac{1}{(k^2-M^2)((p-k)^2-M^2)}
\nonumber \\
           &=& \frac{1}{(4 \pi)^2}(\frac{5}{2}p^2 -4M^2)\int_{0}^{1} d\alpha
\frac{1}{\alpha(1-\alpha) p^2 -M^2} \; ,
\eeq
and
\beq
\label{IHh}
     I_{Hh}(p)&=& \int \frac{d^4k}{(2 \pi)^4}\frac{1}{(k^2-M^2)((p-k)^2-m^2)}
\nonumber \\
    &=&  \frac{1}{(4 \pi)^2}[(-\frac{(m^2-M^2+p^2)^2}{2p^2}+2m^2)
\int_{0}^{1} d\alpha\frac{1}{\alpha(1-\alpha) p^2 -(1-\alpha)m^2-\alpha M^2}]
\nonumber \\
    &&+\frac{5}{2}-\frac{3M^2-m^2}{2p^2}ln(M^2/m^2) \; .
\eeq

  Carrying out the integrals in Eqs(\ref{I1},\ref{I4}), one finds
\beq
\label{I1(p)}
   I1(p)&=& \frac{16 m^2 M^2}{(4\pi)^6}\int_{0}^{1} d\alpha\frac{5(1
-0.8\alpha(1-\alpha))}{2(\alpha(1-\alpha))^2} \nonumber \\
  &&[\frac{m^2-3M^2\alpha(1-\alpha)}{2 \alpha(1-\alpha)}ln(M^2/m^2)
I_{Hh}(p)_{m=0} +\frac{m^2+M^2\alpha(1-\alpha)}{\beta(1-\beta)\alpha(1-\alpha)}
I_{Hh}(p)_{ma^2}\\
  &&-(\frac{m^2}{2(\beta(1-\beta))^2}+\frac{m^2+\alpha(1-\alpha) M^2}
{2 m^2\alpha(1-\alpha)})I_{Hh}(p)_{mb^2} +\frac{m^2+\alpha(1-\alpha) M^2}
{2 m^2\alpha(1-\alpha)}I_{Hh}(p)_{m=0}] \nonumber \; ,
\eeq
with $ma^2=(1-\beta)m_1^2+\beta M^2$ and $mb^2=m^2/(\beta(1-\beta))^2$; and
\beq
\label{I4(p)}
 I4(p)&=&-\frac{32}{(4\pi)^2}m^2\int_{0}^{1}\frac{d\alpha}
{\alpha(1-\alpha)^2}(5\alpha(1-\alpha)-7/2)\int_{0}^{1}\frac{d\gamma}
{1-\gamma} \nonumber \\
 &&[M^2\frac{5m^2-7\bar{m}_2^2/4}{\gamma}I_{Hh}(p)_{\bar{m}_2^2}
+(p^2-m^2)(m_4^2-m_3^2)(2-3\gamma)I_{Hh}(p)_{m_4^2}] \; ,
\eeq
with $m_3^2=(M^2+(1-2\gamma)m_1^2)/(\gamma(1-\gamma))$, $m_4^2=
(M^2+(\gamma-1)m_1^2)/(\gamma(1-\gamma))$, and  $\bar{m}_2^2=
m^2/(\alpha(1-\alpha))$.

Taking the Borel transform one obtains (after removing common factors)
\beq
\label{BI1}
   {\cal B} I1(p)&=&-5.0 \int_{0}^{1} d\alpha \int_{0}^{1} 
d\beta (1-0.8\alpha(1-\alpha))[((m^2-3\alpha(1-\alpha)M^2)ln(M^2/m^2)
\nonumber \\
&&+(m^2+\alpha(1-\alpha)M^2)/(2(1-\beta)m^2-\beta \alpha(1-\alpha)M^2)) 
 \nonumber \\
 &&(M^2(1-\frac{\alpha}{2}-\frac{\beta(1-\beta)}{2})e^{-\alpha M^2/
\beta(1-\beta) M_B^2} \nonumber \\
&& +\int_{0}^{1} d\gamma(1/2)[\frac{(1+\beta)m^2+(1-\alpha)(\alpha+\beta)M^2}
{\bar{M}^2}-\frac{(m^2+\alpha(1-\alpha)M^2)\alpha(1-\alpha)}{(1-\beta)m^2-
\beta \alpha(1-\alpha)M^2}) \nonumber \\
&& (3m^2-M^2+((1-\beta)m^2-\alpha(1-\alpha)M^2)/2)+
\frac{(m^2-M^2)\gamma (1-\gamma)}{2((1-\beta)m^2-\alpha(1-\alpha)M^2)}] 
\nonumber\\
&& e^{-((1-\beta)m^2-\alpha(1-\alpha)M^2)/[\alpha(1-\alpha)
\gamma(1-\gamma)M_B^2]}
\eeq
\beq
\label{BI4}
{\cal B} I4(p)&=&-\int_{0}^{1} d\alpha \int_{0}^{1} 
d\gamma \int_{0}^{1} (5\alpha(1-\alpha)-7/2)[(5\alpha(1-\alpha)\gamma(1-\gamma)
-7/4)m^2 \nonumber \\
&&(m^2-M^2\alpha(1-\alpha)\gamma(1-\gamma)+\frac{((m^2-M^2\alpha(1-\alpha)
\gamma(1-\gamma))^2\lambda(1-\lambda)}{2((1-\lambda)m^2+\lambda\alpha(1-\alpha)
\gamma(1-\gamma)M^2)} \nonumber \\
&& +\frac{1}{2}((1-\lambda)m^2-\lambda\alpha(1-\alpha)\gamma(1-\gamma)M^2))
 \nonumber \\
&& e^{-((1-\lambda)m^2+\lambda\alpha(1-\alpha)\gamma(1-\gamma)M^2)/
\alpha(1-\alpha)\gamma(1-\gamma)\lambda(1-\lambda)M_B^2} \nonumber \\
&&-((3(\gamma-1)m^2+\alpha(1-\alpha)(1-\gamma(1-\gamma))-\frac{1}{2}\bar{M}^2)
\bar{M}^2-\frac{1}{2}((\gamma-1)m^2+\nonumber \\
&&\alpha(1-\alpha)(1-\gamma(1-\gamma))M^2)^2 e^{-\bar{M}^2/[
\alpha(1-\alpha)\gamma(1-\gamma)\lambda(1-\lambda)M_B^2]} 
\eeq

\beq
\label{BPi}
      B\Pi^Z(p)(p)&=& \Pi^Z(M_B)={\cal B} I1(p)+{\cal B} I4(p) \; .
\eeq

 The correlator as a function of the Borel mass, $\Pi^Z(M_B)$  is shown in 
Fig. 2 below.  Note that the result for the mass is given by $M_B$ at the 
minimum in the plot of $\Pi^Z(M_B)$, and the error by the shape of the plot 
near the minimum\cite{svz79}, as is discussed in detail in Ref\cite{lsk09}.
\vspace{-1cm}

\begin{figure}[ht]
\begin{center}
\epsfig{file=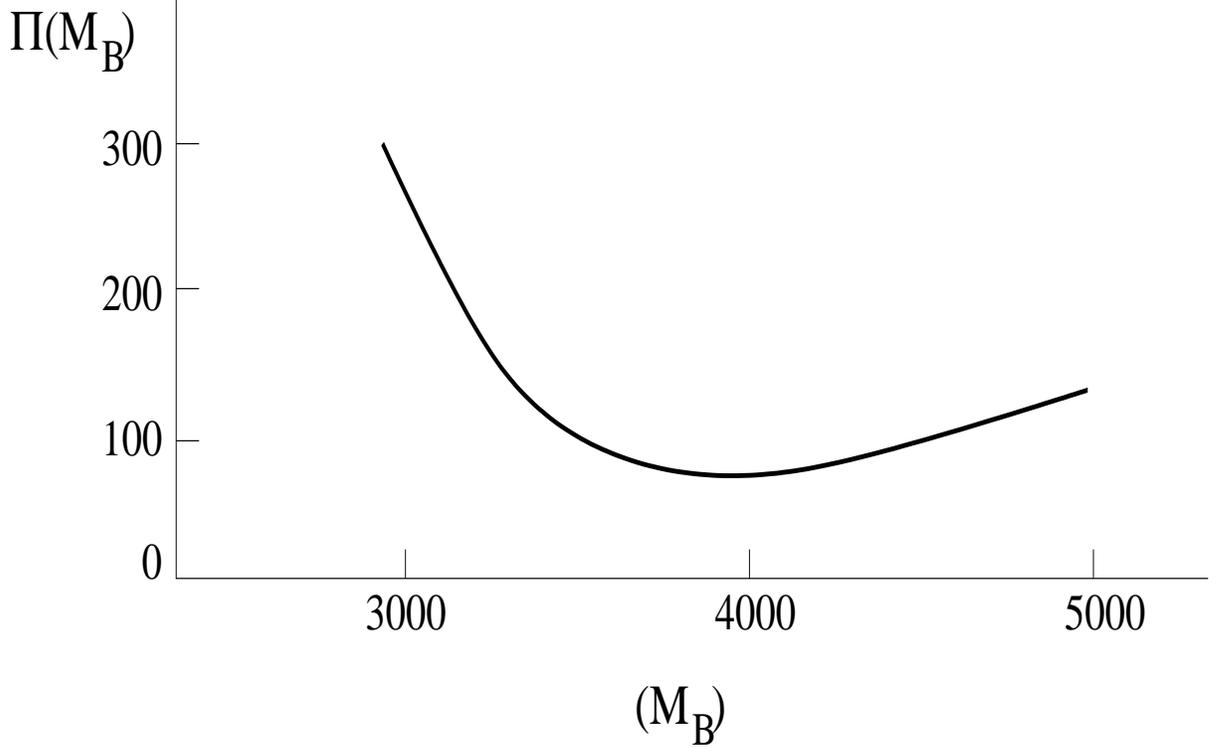,height=10cm,width=16cm}
\caption{$\Pi(M_B)$ is the 4-quark correlator,a function of the
Borel Mass $M_B$, in units of MeV}
\label{Fig.2}
\end{center}
\end{figure}
\newpage
\section{Results and Conclusions}

  From Figure 2, the mass of the $|c\bar{d}\bar{c}u>$ state is 
$3900 \pm 200$ MeV, in agreement with the state found in the recent 
BESIII\cite{BESIII13} and
Belle\cite{Belle13} experiments. From this we conclude that the conjecture
of these two collaboretions is correct, that the $Z_c(3900)$ is a 
tetra-quark state. For decades experimentalists and theorists have 
attempted to find tetra-quark states, so this is a very important discovery.

Our plans for future research include estimates of the decay probabilities of
our tetra-quark thery of the $Z_c(3900)$ to compare with experimental results,
which requires a three-point correlator. 

\vspace{3mm}
\Large{{\bf Acknowledgements}}\\
\normalsize
This work was supported in part by a grant from the Pittsburgh Foundation.
LSK thanks LANL experimentalists for suggesting that the BESIII and Belle
experiments could be an important discovery in high energy particle physics, 
and for a discussion with Prof. Eric Swanson.

\end{document}